\documentstyle[12pt]{article}

\author{Hao Chen\\
Department of Mathematics\\
Zhongshan University\\
Guangzhou,Guangdong 510275\\
People's Republic of China}
\title{Constraints on the mixing of states on bipartite quantum systems}
\date{September,2002}

\begin{document}

\maketitle
\begin{abstract}
We give necessary conditions for the mixing of states on bipartite quantum systems, which are independent of the eigenvalues of these mixed states and based on the algebraic-geometric invariants introduced in [1]. These are further constraints except the previously known constraints based on majorization of eigenvalues ([2,3,4]). One implication of our results is that for some special mixed states, only mixed states in a measure zero set can be used to mix to get them. As indicated in [1] and [5] for many physical problems in composite quantum systems the majorization of some numerical invariants such as eigenvalues is not sufficient. 

\end{abstract}

The general problem of quantum operations was considerd in [6],[7],[8], essentially new aspects from the view of quantum information processing was added recently in [9],[10],[11],[12],[13]. All possible physical operations on quantum systems can be divided into (1)state-to-ensemble operations and (2)ensemble-to-state operations(mixing), where ensemble is a set of mixed states $\{\rho_i\}$ with ascribed possibilities $\{p_i\}$. The class (2) can be described by taking convex combination\\

$$
\begin{array}{ccccc}
\{p_i,\rho_i\} \rightarrow \rho_{out}=\Sigma p_i \rho_i\\
\end{array}
(1)
$$

In the context of quantum information processing, the action of mixing corresponds to erasure of informations concerning identity of a member of ensemble. An important problem in quantum mechnics is the following mixing problem: given a $\rho$, characterize the ensembles $\{p_i,\rho_i\}$ such that $\rho=\Sigma p_i \rho_i$ ([14],[3]).

In uni-partite case there are some previously known constraints on $\{p_i,\rho_i\}$ for a given mixed state $\rho$ based on majorization of eigenvalues of $\rho$, $\rho_i$ and $p_i$ (see [2],[3],[4],[15],[16]). We first recall the notation of majorization ([2],[3],[4]). For two given vectors of real numbers $r$ and $s$, we re-order the components of $r$ and $s$ into decreasing order, writing for example $r^{\downarrow}=(r_1^{\downarrow},...,r_d^{\downarrow})$ for the vector whose components are the same as those of $r$, but in decreasing order. We say $r$ is majorized by $s$ , written $r \prec s$ if \\

$$
\begin{array}{ccccccccc}
r_1^{\downarrow} \leq s_1^{\downarrow}\\
r_1^{\downarrow}+r_2^{\downarrow} \leq s_1^{\downarrow}+s_2^{\downarrow}\\
\cdots\\
r_1^{\downarrow}+...+r_{d-1}^{\downarrow} \leq s_1^{\downarrow}+...+s_{d-1}^{\downarrow}\\
r_1^{\downarrow}+...+r_d^{\downarrow} = s_1^{\downarrow}+...+s_d^{\downarrow}\\
\end{array}
(2)
$$

Then we can recall the previously known constraints of the mixing problem in uni-partite case.\\

{\bf Theorem 1 ([15],[16],[3],[4]).} {\em Suppose $\rho$ is a mixed state. Let $\{p_i\}$ ba a probility distribution. Then there exist nomarlized quantum states $|\phi_i>$ such that $\rho=\Sigma_i p_i |\phi_i><\phi_i|$ if and only if $(p_i) \prec \lambda (\rho)$, where $\lambda(\rho)$ is the vector of eigenvalues of $\rho$.} \\

{\bf Theorem 2 ([4]).} {\em Let $\rho$ be a mixed state, $\{p_j\}$ a probility distribution and $\{\rho_j\}$  mixed states such at $\rho=\Sigma_j p_j \rho_j$. Then the following constraints must be obeyed: $\lambda (\rho) \prec \Sigma_j p_j \lambda (\rho_j)$, where $\lambda ()$ denotes the vector of eigenvalues.}\\

We now consider the bipartite case. Let $\rho$ and $\rho_i$'s be mixed states on $H_A^m \otimes H_B^n$ and $\rho =\Sigma p_i \rho_i$ where $p_i >0$. It is clear that we can take partial traces of both sides and get some constraints on the eigenvalues of $tr_A(\rho)$ and $tr_A(\rho_i)$ from Theorem 2 (respectively $tr_B(\rho)$ and $tr_B(\rho_i)$). However there are other constraints for the mixing problem of bipartite mixed states  related to Schmidt numbers. First we need to recall the concept of Schmidt numbers of bipartite mixed states introduced in [17]. For a bipartite mixed state $\rho$, it has Schmidt number $k$ if and only if for any decomposition $\rho=\Sigma_i p_i |v_i><v_i|$ for positive real numbers $p_i$'s and pure states $|v_i>$'s, at least one of the pure states $|v_i>$'s has Schmidt rank at least $k$, and there exists such a decomposition with all pure states $|v_i>$'s Schmidt rank at most $k$. It is clear that the mixed states are entangled if their Schmidt numbers are bigger than 1. It is proved ([17]) that Schmidt number is entanglement monotone, ie., they cannot increase under local quantum operations and classical communication . So we can naturally think Schmidt numbers of mixed states as a measure of their entanglement. \\

  We have the following necessary condition of mixing problem in bipartite case, which is independent of eigenvalues.\\

{\bf Proposition 1.} {\em Let $\rho$ be a mixed state on a bipartite system with Schmidt number $r$. Suppose  $\rho=\Sigma p_i |\phi_i><\phi_i|$ where $|\phi_i>$'s are pure states and $p_i$'s are positive. Then one of $|\phi_i>$ has Schmidt rank at least $r$. In the case that $r \geq 2$, ie., $\rho$ is entangled, then at least one of $|\phi_i>$ is entangled.}\\

From this observation it is clear that the mixing problem in bipartite case is more complicated than the uni-partite case and cannot be characterized by only using eigenvalues. In [1] algebraic-geometric invariants of bipartite mixed states under local unitary transformations are introduced and it is showed that these invariants can help us to understand the physical problems such as separability of mixed states and simulation of Hamiltonians ([1],[5]). In this paper we give some necessary conditions of the mixing problem in bipartite case in the term of only these algebraic-geometric invariants (thus independent of eigenvalues). First we need to recall the definition of these invariants. For any bipartite mixed states $\rho$ on $H_A^m \otimes H_B^n$ , we want to understand it by measuring it with separable pure states, ie., we consider the $<\phi_1 \otimes \phi_2
|\rho|\phi_1 \otimes \phi_2>$ for any pure states $\phi_1 \in H_A^m$ and $\phi_2 \in H_B^n$. For any fixed $\phi_1 \in P(H_A^m)$, where $P(H_A^m)$ is
the projective space of all pure states in $H_A^m$, $<\phi_1 \otimes \phi_2
|\rho|\phi_1 \otimes \phi_2>$ is a Hermitian bilinear form on $H_B^n$,
denoted by $<\phi_1|\rho|\phi_1>$ . We consider the {\em degenerating locus }
of this bilinear form, ie., $V_A^k(\rho)=\{\phi_1 \in P(H_A^m): rank
(<\phi_1|\rho|\phi_1>) \leq k\}$ for $k=0,1,...,n-1$. We can use the
coordinate form of this formalism. Let $\{|11>,...,|1n>,...,|m1>,...,|mn>\}$
be the standard orthogonal base of $h_A^m \otimes H_B^n$ and $\rho$ be an
arbitrary mixed states. We represent the matrix of $\rho$ in the base $\{|11>,...|1n>,...,|m1>,...,|mn>\}$, and consider $\rho$ as a blocked matrix 
$\rho=(\rho_{ij})_{1 \leq i \leq m, 1 \leq j \leq m}$ with each block $\rho_{ij}$ a $n \times n$ matrix corresponding to the $|i1>,...,|in>$ rows
and the $|j1>,...,|jn>$ columns. For any pure state 
$\phi_1=r_1|1>+...+r_m|m> \in P(H_A^m)$ the matrix of the Hermitian linear
form $<\phi_1|\rho|\phi_1>$ with the base $|1>,...,|n>$ is $\Sigma_{i,j}
r_ir_j^{*} \rho_{ij}$. Thus the ``degenerating locus'' is actually as
follows.\\
$$
\begin{array}{ccccc}
V_{A}^k(\rho)=\{(r_1,...,r_m)\in CP^{m-1}:rank( \Sigma_{i,j}r_ir_j^{*}
\rho_{ij}) \leq k\} &  &  &  & 
\end{array}
$$
for $k=0,1,...,n-1$. Similarly $V_{B}^k (\rho) \subseteq CP^{n-1}$ can be
defined. Here * means the conjugate of complex numbers. It is known from
Theorem 1 and 2 of [1] that these sets are algebraic sets (zero locus of
several multi-variable polynomials, see [18]) and they are invariants under
local unitary transformations depending only on the eigenvectors of $\rho$.
Actually these algebraic sets can be computed easily as follows.\\

Let $\{|11>,...,|1n>,...,|m1>,...,|mn>\}$ be the standard orthogonal base of 
$H_A^m \otimes H_B^n$ as above and $\rho= \Sigma_{l=1}^{t} p_l |v_l><v_l|$
be any given representation of $\rho$ as a convex combination of projections
with $p_1,...,p_t >0$. Suppose $v_l=\Sigma_{i,j=1}^{m,n} a_{ijl} |ij>$ , $A=(a_{ijl})_{1\leq i \leq m, 1 \leq j \leq n, 1 \leq l \leq t}$ is the $mn
\times t$ matrix. Then it is clear that the matrix representation of $\rho$
with the base $\{|11>,...,|1n>,...,|m1>,...,|mn>\}$ is $AP(A^{*})^{\tau}$,
where $P$ is the diagonal matrix with diagonal entries $p_1,...,p_t$. We may
consider the $mn\times t$ matrix $A$ as a $m\times 1$ blocked matrix with
each block $A_w$, where $w=1,...,m$, a $n\times t$ matrix corresponding to 
$\{|w1>,...,|wn>\}$. Then $V_A^k(\rho)$ is just the algebraic set in $CP^{m-1}
$ as the zero locus of the determinants of all $(k+1) \times (k+1)$
submatrices of $\Sigma_i^m r_i A_i$.\\

Now we can state the main results of this paper.\\

{\bf Theorem 3.} {\em Let $\rho, \rho_i$  be bipartite mixed states on $H_A^m \otimes H_B^n$ and $p_i$'s be  positive real numbers. Suppose $\rho=\Sigma p_i \rho_i$. Then $V_A^j(\rho) \subset V_A^j(\rho_i)$ and $V_B^j(\rho) \subset V_B^j(\rho_i)$ for any possible indices $i$ and $j$.}\\

{\bf Proof.} Suppose $\rho=\Sigma_i^h p_i \rho_i$. For any $\rho_i$, if $\rho_i=\Sigma_t^{s_i} q_t^i |v_t^i><v_t^i|$ where $q_t^i >0$. Then $\rho=\Sigma_{it} p_iq_t^i |v_t^i><v_t^i|$. From the definition we can compute algebraic-geometric invariants $V_A^k(\rho)$ of $\rho$ from all vectors $v_t^i$'s, where $i=1,...,h$ and $t=1,...,s_i$, and compute $V_A^k(\rho')$ from vectors $v_t^i$, where $t=1,...,s_i$. Thus the matrix $\Sigma_j ^m r_j A_j$ of $\rho_i$ is a submatrix of the corresponding matrix of $\rho$ and the conclusion is proved.\\

Because algebraic-geometric invariants of bipartite mixed states are independent of eigenvalues of the states, thus the constriants in Theorem 3 is essentially different with previously known constraints in Theorem 1 and 2.\\

We need the following result. For pure states $\rho=|v><v|$ on $H_A^m \otimes H_B^n$ with $m \leq n$, we can compute its algebraic-geometric invariants  from its Schmidt decomposition $v= \Sigma_{i=1}^d a_i e_i \otimes e'_i$, where $e_1,...,e_m$ (resp., $e'_1,...,e'_n$) is a orthogonal base of $H_A^m$ (resp. $H_B^n$). It is clear that $V_A^0(\rho) =\{(r_1,...,r_m) \in CP^{m-1}: (a_1r_1,...,a_d r_d,0,...,0)^{\tau}=0\}$. Thus we have\\

{\bf Proposition 2.} {\em For the pure state $\rho=|v><v|$, $d=m$ if and only if $V_A^0(\rho)=\emptyset$ and $d=m-1-dim(V_A^0(\rho))$ if $d \leq m-1$.}\\

From Proposition 2 the following result is an easy implication of Theorem 3.\\

{\bf Corollary 1.} {\em Let $\rho$ be a mixed state on $H_A^m \otimes H_B^n$, $|\phi_i>$'s be pure states and $p_i$'s be positive real numbers. Suppose  $\rho =\Sigma p_i |\phi_i><\phi_i|$ . Then the following constraints hold.\\

$$
\begin{array}{ccccccccccccccc}
k(|\phi_i>) \leq m-1-dim V_A^0(\rho)\\
k(|\phi_i>) \leq n-1-dim V_B^0(\rho)
\end{array}
(1)
$$

, where $k(|\phi_i>)$ is the Schmidt rank of the pure state $|\phi_i>$. Here $dim V_A^0(\rho)$ or $dim V_B^0(\rho)$ is $-1$ if it is empty set.}\\

The following example is a simple application of Corollary 1.\\

{\bf Example 1.} Let $\rho=\frac{1}{2}(|\phi_1><\phi_1|+|\phi_2><\phi_2|)$ where\\

$$
\begin{array}{cccccccccccc}
\phi_1=\frac{1}{2}(|11>+|12>+|21>+|22>)\\
\phi_2=\frac{1}{\sqrt{2}}(|11>+|21>)
\end{array}
(2)
$$

on $H_A^2 \otimes H_B^2$ be a separable rank 2 mixed states. If there are some pure states $|\psi_i>$'s such that $\rho=\Sigma p_i |\psi_i><\psi_i|$ where $p_i >0$. From Proposition 1 we cannot say anything about $|\psi_i>$'s since the Schmidt number of $\rho$ is 1. However we can compute that $V_A^0(\rho) \subset CP^1$ is one point $(1:-1)$. Thus from Corollary 1 it is clear that the Schmidt rank of each $|\psi_i>$ has to be 1, ie., each $|\psi_i>$ is separable.\\

This can  be generalized to the following result, which is direct from Corollary 1.\\

{\bf Corollary 2.} {\em If $\rho$ is a mixed state on $H_A^m  \otimes H_B^n$ with $dim V_A^0(\rho)=m-2$ or $dim V_B^0(\rho)=n-2$. Suppose $\rho=\Sigma p_i |\phi_i><\phi_i|$ where $p_i >0$. Then each $|\phi_i>$ is a sepaprable pure state.}\\

The following example is an application of Theorem 3.\\

{\bf Example 2.} Let $\rho=\frac{1}{4}\Sigma_{i=1}^4 |\phi_i><\phi_i|$ be a rank 4 mixed state on $H_A^3 \otimes H_B^3$, where,\\

$$
\begin{array}{ccccccccccccccccc}
\phi_1=|11>\\
\phi_2=\frac{1}{\sqrt{2}}(|21>+|32>)\\
\phi_3=\frac{1}{\sqrt{2}}(|12>+|33>)\\
\phi_4=|33>
\end{array}
(3)
$$

and $\rho'=\frac{1}{3}\Sigma_{i=1}^3|\psi_1><\psi_i|$ be a rank 3 mixed state on $H_A^3 \otimes H_B^3$, where,\\

$$
\begin{array}{ccccccccccccccc}
\psi_1=\frac{1}{\sqrt{3}}(|11>+|22>+|33>)\\ 
\psi_2=\frac{1}{\sqrt{3}}(|12>+|23>+|31>)\\ 
\psi_3=\frac{1}{\sqrt{3}}(|13>+|21>+|32>)
\end{array}
(4)
$$

It is easy to compute  that $\Sigma_i r_i A_i$ for $\rho$ is the following matrix\\
$$
\left(
\begin{array}{ccccccccccccc}
r_1&r_2&0&0\\
0&r_3&r_1&0\\
0&0&r_2&r_3\\
\end{array}
\right)
(5)
$$

and thus the line in $CP^2$ defined by $r_1=0$ is in $V_A^2(\rho)$. However it is easy to compute that $\Sigma_i r_i A_i$ for $\rho'$ is the following matrix.\\

$$
\left(
\begin{array}{ccccccccccccc}
r_1&r_2&r_3\\
r_2&r_3&r_1\\
r_3&r_1&r_2\\
\end{array}
\right)
(6)
$$

We can check that for some points $(0:r_2:r_3)$ in $CP^2$, the rank of (6) is 3. Thus $V_A^2(\rho)$ is not contained in $V_A^2(\rho')$ and from Theorem 3, there is no positive reals $p',p_i's$ and mixed states $\rho_i's$ such that $\rho=p'\rho'+\Sigma_i p_i \rho_i$.\\

From Corollary 1 and Proposition 2 we have the following result.\\

{\bf Corollary 3.}{\em If $\rho$ is a mixed state on $H_A \otimes H_B^n$ with $V_A^0(\rho) \neq \emptyset$ or $ V_B^0(\rho) \neq \emptyset$. Suppose  $\rho=\Sigma p_i |\phi_i><\phi_i|$ where $p_i >0$. Then the Schmidt rank of each $|\phi_i>$ cannot be $min\{m,n\}$.}\\

{\bf Example 3.}  Let $\rho=\frac{1}{4}\Sigma_{i=1}^4 |\phi_i><\phi_i|$ be a rank 4 mixed state on $H_A^3 \otimes H_B^3$, where,\\

$$
\begin{array}{ccccccccccccccccc}
\phi_1=|11>\\
\phi_2=\frac{1}{\sqrt{5}}(|12>+|21>+|22>+|31>+|32>)\\
\phi_3=\frac{1}{\sqrt{3}}(|13>+|22>+|32>)\\
\phi_4=\frac{1}{\sqrt{2}}(|23>+|33>)
\end{array}
(7)
$$

It is esay to compute that $\Sigma_i r_i A_i$ for $\rho$ is the following matrix\\

$$
\left(
\begin{array}{ccccccccccccc}
r_1&r_2+r_3&0&0\\
0&r_1+r_2+r_3&r_2+r_3&0\\
0&0&r_1&r_2+r_3\\
\end{array}
\right)
(8)
$$

Thus $V_A^0(\rho)$ is the one point $(0:1:-1)$ in $CP^2$, not empty, thus only pure states with their Schmidt rank smaller than 3 can be used to mix to get $\rho$.\\

Here we should note that in the set of all pure states $\{|\phi>=\Sigma a_{ij}|ij>\}$, since the Schmidt rank of $|\phi>$ is the rank of the matrix $A=(a_{ij})_{1\leq i \leq m, 1\leq j \leq n}$, the set of all pure states with their Schmidt ranks smaller than $min\{m,n\}$ is a measure zero set definded by $det(A)=0$. Thus Corollary 2 indicates that for mixed states $\rho$ with $V_A^0(\rho)$ or $V_B^0(\rho)$ not empty, only pure states in a measure zero set can be used to mix to get $\rho$. Actually this is quite common phenomenon even for mixed states as illustrated in the following Theorem 4.\\

{\bf Theorem 4.} {\em Let $r$ and $t$ be non-negative integers satisfying $1 \leq r \leq mn$ and $t \geq \frac{m+r+\sqrt{(m+r)^2-4mr+4m}}{2}$ or $t \leq \frac{m+r+\sqrt{(m+r)^2-4mr+4m}}{2}$, $\rho$ be a mixed state on $H_A^m \otimes H_B^n$ with $rank(\rho) \geq r$. If $V_A^t(\rho) \neq \emptyset$, then the set of rank $r$ mixed states $\rho'$ for which there exist positive reals $p'$ and $p_i's$ and mixed states $\rho_i's$  such that $\rho=p'\rho' +\Sigma p_i \rho_i$ is a measure zero set.}\\

{\bf Example 4.}  Let $\rho=\frac{1}{5}\Sigma_{i=1}^5 |\phi_i><\phi_i|$ be a rank 5 mixed state on $H_A^4 \otimes H_B^4$, where,\\

$$
\begin{array}{ccccccccccccccccc}
\phi_1=|11>\\
\phi_2=\frac{1}{\sqrt{2}}(|12>+|21>)\\
\phi_3=\frac{1}{\sqrt{3}}(|32>+|33>+|41>)\\
\phi_4=\frac{1}{2}(|31>+|34>+|42>+|43>)\\
\phi_5=\frac{1}{\sqrt{3}}(|12>+|14>+|23>)
\end{array}
(9)
$$

It is esay to compute that $\Sigma_i r_i A_i$ for $\rho$ is the following matrix\\

$$
\left(
\begin{array}{ccccccccccccc}
r_1&r_2&r_4&r_3&0\\
0&r_1&r_3&r_4&r_1\\
0&0&r_3&r_4&r_2\\
0&0&0&r_3&r_1
\end{array}
\right)
(10)
$$

We take $r=4$ and $t=2$. It is clear that the line in $CP^3$ defined by $r_1=r_2=0$ is in $V_A^2(\rho)$. Thus from Theorem 4, the set of rank 4 mixed states $\rho'$ on $H_A^4 \otimes H_B^4$ for which there exist positive reals $p'$ and $p_i's$ and mixed states $\rho_i's$ such that $\rho= p'\rho' + \Sigma p_i \rho_i$ is a measure zero set in the set of all rank 4 mixed states.\\

For the purpose to prove Theorem 4, we need to recall a well-known result in
the theory of determinantal varieties (see Proposition in p.67 of [19]). Let 
$M(m,n)=\{(x_{ij}): 1\leq i \leq m, 1 \leq j \leq n\}$ (isomorphic to 
$CP^{mn-1}$) be the projective space of all $m \times n$ matrices. For a
integer $0 \leq k \leq min\{m,n\}$, $M(m,n)_k$ is defined as the locus $%
\{A=(x_{ij}) \in M(m,n): rank(A) \leq k\}$. $M(m,n)_k$ is called generic
determinantal varieties.\\

{\bf Proposition 3.} {\em $M(m,n)_k$ is an irreducible algebriac subvariety
of $M(m,n)$ of codimension $(m-k)(n-k)$.}\newline

We describe the basic idea of Proposition 3. Since all entries in the $m\times n$ matrix are indeterminants
, we can suppose that the 1st $k \times k$ submatrix is nonsingular and the 
remaining $m-k$ columns ($n-k$ rows) are linear dependent on the 1st $k$ columns
($k$ rows). This condition implies that the determinants of all $(m-k)(n-k)$ $(k+1) \times (k+1)$ submatrices containing
1st $k \times k$ submatrix are zero, ie., we have $(m-k)(n-k)$ (independent) algebraic equations to define $M(m,n)_k$, thus the conclusion of Proposition 3 is valid.\\  

{\bf Proof of Theorem 4.} From Theorem 3, we know that for $\rho'$, if  there exist positive reals $p'$ and $p_i's$ and mixed states $\rho_i's$  such that $\rho=p'\rho' +\Sigma p_i \rho_i$, then $V_A^t(\rho')$ is not empty. Thus we only need to prove that $V_A^t(\rho')$ of generic (generic=outside a measure zero set) rank $r$ mixed states on $H_A^m \otimes H_B^n$ is empty. From Propositin 3 the codimention of $V_A^t(\rho')$ of generic rank $r$ mixed states on $H_A^m \otimes H_B^n$ is $(m-t)(r-t)$. From the condition of $t$ in Theorem 4, $(m-t)(r-t) \geq m$, thus $V_A^t(\rho')$ in $CP^{m-1}$ is empty. The conclusion is proved.\\

In conculsion, we have proved necessary conditions for the mixing of states on bipartite quantum systems , which are independent of eigenvalues of states and essentially different with previously known constraints. We also proved that for some special bipartite mixed states, only special mixed states in a measure zero set can be used to mix to get them. These results indicate that for the mixing problem in bipartite case, except previously known constraints based on majorization of eigenvalues, there are other very strong constraints described by algebraic-geometric invariants. \\

The author acknowledges the support from NNSF China, Information Science Division, grant 69972049.\\

e-mail: chenhao1964cn@yahoo.com.cn\\

\begin{center}
REFERENCES
\end{center}

1.H. Chen, quant-ph/01008093\\

2.M.A.Nielsen, Phys. Rev. A, 63(02), 022144,2000\\

3.M.A.Nielsen, Phys.Rev. A, 62:052308, 2000\\

4.M.A.Nielsen and G.Vidal, Quantum Information and Computation, Vol.1, No.1: 76-92, 2001\\

5.H.Chen, quant-ph/0109115\\

6.K Kraus, States, Effects and Operations (Springer, Berlin, 1983)\\

7.E.B.Davis, Quantum Theory of Open Systems, Academic, London, 1976\\

8.P.Bush, M.Grabowski and Lahti, Operational Quantum Physics, Springer, 1997\\

9.C.H.Bennett,D.P.DiVincenzo,J.Smolin and W.K.Wooters, Phys.Rev.A 54, 3814(1997)\\

10.G.Vidal, J.Mod.Opt. 47, 355(2000)\\

11.A.Peres, Phys.Rev. A 61(2000) 022116\\

12.P.Horodecki, Ph. D. Thesis , Politechnika Gdanska, Gdanska 1999\\

13.E.M.Rains, Phys. Rev. A 60, 173 (1999)\\

14.M.Horodecki,Quantum Information and Computation, Vol.1, No.1:3-26, 2001\\

15.A.Uhlmann, Rep. Math. Phys. 1(2):147-159, 1970\\

16.L.P.Hughston, R.Jozsa and W.K.Wooters, Phys. Lett. A, 183:14-18, 1983\\

17.B.M.Terhal and P.Horodecki, Phys. Rev. A R040301, 61(2000)\\

18.J.Harris, Algebraic Geometry, A First course, GTM 133, Springer-Verlag
1992\\

19.E.Arbarello, M.Cornalba, P.A.Griffiths and J.Harris, Geometry of
algebraic curves,Volume I, Springer-Verlag, 1985, Chapter II''Determinantal
Varieties''\\

\end{document}